\documentclass[12pt]{iopart}

\usepackage{iopams}
\usepackage{graphicx}

\newtheorem{conjecture}{Conjecture}
\newtheorem{theorem}{Theorem}

\begin{document}

\title[Saari's homographic conjecture in  planar three-body problem]
{Saari's homographic conjecture
for general masses
in  planar three-body problem
under  Newton potential and a strong force potential}

\author{
Toshiaki Fujiwara$^1$,
Hiroshi Fukuda$^2$,
Hiroshi Ozaki$^3$\\
and Tetsuya Taniguchi$^4$
}
\address{
$^{1,2,4}$ College of Liberal Arts and Sciences, Kitasato University,
1-15-1 Kitasato, Sagamihara, Kanagawa 252-0329, Japan
}
\address{
$^3$ General Education Program Center, Tokai University, Shimizu Campus,
3-20-1, Orido, Shimizu, Shizuoka 424-8610, Japan
}

\eads{
$^{1}$fujiwara@kitasato-u.ac.jp,
$^{2}$fukuda@kitasato-u.ac.jp,
$^{3}$ozaki@tokai-u.jp,
$^{4}$tetsuya@kitasato-u.ac.jp
}

\begin{abstract}
Saari's homographic conjecture claims that,
in the $N$-body problem
under the homogeneous potential, $U=\alpha^{-1}\sum m_i m_j/r_{ij}^\alpha$ for $\alpha\ne 0$,
%
a motion having constant configurational measure
 $\mu=I^{\alpha/2}U$ is homographic,
 where $I$ represents the moment of inertia
 defined by $I=\sum m_i m_j r_{ij}^2/\sum m_k$, 
 $m_i$ the mass, and $r_{ij}$  the distance between particles.

We prove this conjecture for general masses $m_k>0$
in the planar three-body problem
under  Newton potential ($\alpha=1$) and
a strong force potential ($\alpha=2$).
\end{abstract}


\section{
\label{sec:Intro}
Saari's homographic conjecture and construction of this paper}
Consider the $N$-body problem described by the Lagrangian
$L=K/2+U$.
The $K$ here represents twice of the kinetic energy
\begin{equation}
\label{defOfK}
K=\sum_{k=1,2,\dots,N} m_k \left|\frac{d\bi{q}_k}{dt} \right|^2,
\end{equation}
and the $U$ represents homogeneous potential function
%
\begin{equation}
\label{defOfPotential}
U=
\cases{
	\frac{1}{\alpha}
		\sum_{1\le i < j \le N} \frac{m_i m_j}{|\bi{q}_i-\bi{q}_j|^\alpha}
		& for $\alpha\ne 0$,\\
	-\sum_{1\le i < j \le N} m_i m_j \log |\bi{q}_i-\bi{q}_j|
		& for $\alpha=0$.
}
\end{equation}
Here, $m_k$ and $\bi{q}_k \in \mathbb{R}^3$ represent
the mass and the position vector of the point particle $k=1,2,\dots,N$.
The real parameter $\alpha$ represents the power of
 mutual distance of point particles.
The values $\alpha=1$, $2$, and $-2$ give  Newton potential,
a strong force potential,
and a harmonic oscillator potential respectively.
Although the logarithmic potential is not invariant under a scale transformation,
we take this potential for $\alpha=0$.
This is because
a scale transformation for this potential
adds a constant term that has no effect for the
equations of motion.
Indeed, the definition \eref{defOfPotential} makes the equations of motion 
\begin{equation*}
  m_k \frac{d^2 \bi{q}_k}{dt^2}
  =\sum_{i \ne k} \frac{m_i m_k (\bi{q}_i-\bi{q}_k)}{|\bi{q}_i-\bi{q}_k|^{\alpha+2}}
\end{equation*}
for all $\alpha \in \mathbb{R}$.
%
The moment of inertia $I$ is defined as follows,
\begin{equation}
\label{defOfMomentOfInertia}
I=\left(\sum_{1\le i<j \le N} m_i m_j |\bi{q}_i-\bi{q}_j|^2\right)
	\left(\sum_{k=1,2,\dots,N} m_k\right)^{-1}.
\end{equation}
The configurational measure $\mu$ is defined as a
scale invariant product of $I$ and $U$,
as follows,
\begin{equation}
\mu=
\cases{
	\alpha I^{\alpha/2}U & for $\alpha \ne 0$,\\
	-\sum m_i m_j \log \left(|\bi{q}_i-\bi{q}_j|/\sqrt{I}\right) & for $\alpha=0$.
}
\end{equation}

Saari's homographic conjecture, which is the subject of this paper,
may have several expressions.
One expression is the following.
\begin{conjecture}[Saari's homographic conjecture, 2005]
\label{homographicConjecture}
For homogeneous potential with an arbitrary $\alpha$
where the configurational measure $\mu$ is not identically constant,
if a motion has a constant value of $\mu$
then the motion is homographic.
\end{conjecture}
This conjecture consists of two parts.
The first part states that some exceptional cases should be excluded 
and the second part states the body of the conjecture.

The statement for the exceptional cases may need some explanations.
If $\mu$ is identically constant,
in other words, if $\mu$ is constant for any motion,
constancy of $\mu$ obviously give no restriction for the motion.
This will take place when $U$ is proportional to $I^{-\alpha/2}$.
There are two known cases.
Case~1: Harmonic oscillator ($\alpha=-2$), $U=-I \sum m_k/2$.
Therefore, $\mu=-2I^{-1}U=\sum m_k$ is identically constant.
Case~2: Three-body equal-mass rectilinear motion in $\alpha=-4$
\cite{Chenciner1997}.
In this case,
$U=-\sum (x_i -x_j)^4/4=-9 I^2/8$
is an identity
for $x_k \in \mathbb{R}$, $k=1,2,3$.
Therefore, $\mu=-4I^{-2}U=9/2$ is identically constant.
We expect that
there	 are no more  exceptional cases.

A motion is called homographic
if the configuration $\{\bi{q}_k(t)\}$ remains similar to the original configuration $\{\bi{q}_k(0)\}$.
Here, the similarity is defined by scale transformation and rotation.
In other word,
for a homographic motion in planar $N$-body problem,
there exists a complex function $z(t)$, such that
\begin{equation}
\label{defOfHomographic}
q_k(t)=z(t)q_k(0).
\end{equation}
Although the term ``similarity transformation'' usually contains 
parallel transformation and reverse transformation,
we exclude them if we do not explicitly mention otherwise.
The parallel transformation is excluded
because we always consider the centre of mass frame in this paper.
The reverse transformation is excluded
because we are considering a dynamical motion
which  is always continuous in  time.

The converse of the conjecture \ref{homographicConjecture} is obviously true.
This is because  $\mu$ is invariant
under the above similarity transformations.
Therefore, if the motion is homographic then $\mu$ is constant.

The aim of this paper is to prove
Saari's homographic conjecture in planar three-body problem
for general masses with $\alpha=1$, $2$.
Namely, we will prove the following theorem.
\begin{theorem}
For planar three-body problem
with $\alpha=1$ and $2$,
if a motion has constant $\mu$
then the motion is homographic.
\end{theorem}

The construction of this paper is the following.
In the section \ref{sec:Saari'sconjectures}, we give a  history of  Saari's conjecture.
In the section \ref{sec:DynamicalVariables},
we introduce dynamical variables to describe the motion 
of size, rotation and shape.
Then, we obtain the Lagrangian
and  derive the equations of motion for these variables
under the potential $\alpha\ne 0$.
%
In the section \ref{sec:NecessaryCondition},
non-homographic motion with constant $\mu$
will be assumed to  exist.
Then, we will obtain a necessary condition for such motion
to be compatible to the equations of motion.
In the section \ref{sec:Proof},
we prove that the necessary condition is not satisfied
for $\alpha=1$, $2$.
This means that
there is no non-homographic motion with constant $\mu$.
This is a proof of  Saari's homographic conjecture
for $\alpha=1$ and $2$.
%
Summary and discussions are given in the section 6.

\section{Saari's conjectures}
\label{sec:Saari'sconjectures}
%
By now, three conjectures are named after ``Saari''.
Donald Saari   stated his conjecture in 1969  in the $N$-body problem under Newton potential
which we would like to call it 
``Saari's original conjecture''.
Then, people extended the original conjecture to general homogeneous potential
which were called  ``generalised Saari's  conjecture''.
Finally, in 2005, Saari extended his conjecture which we  call it 
``Saari's homographic conjecture''.

In this section,
we will describe  each conjecture,
its brief history and current status of 
known exceptions and proofs.
See also \Tref{SummaryForSaarisThreeConjectures}.

\begin{center}
\begin{table}
\centering
\caption{Summary for three Saari's conjectures.
}
\label{SummaryForSaarisThreeConjectures}
\begin{tabular}{p{4.5em}|p{4.5em}|p{10em}|p{12em}}
\hline\hline
conjecture 	& original	& generalised	& homographic\\
\hline
range of $\alpha$& \multicolumn{1}{c|}{$\alpha=1$}& \multicolumn{2}{c}{$\alpha \in \mathbb{R}$} \\
\hline
assumption&\multicolumn{2}{c|}{$I$=const.}&$\mu$=const.\\
\hline
known exceptions & & \multicolumn{2}{p{22em}}
{$\alpha=-2$ and equal-mass rectilinear 3-body in $\alpha=-4$, where $\mu$ is trivially constant.} \\ 
& & $\alpha=2$&
$^\#$\\
\hline
proof& 
3-body in spacial dimension $\ge 2$. & For $\alpha\ne 2$, ``generalised'' is contained in ``homographic''. & 
Collinear $N$-body for any $\alpha$ and \emph{equal-mass}$^\dag$ planar 3-body for $\alpha=1,2$.\\
\hline\hline
\end{tabular}
$^{\#}$Saari's homographic conjecture is expected to be true for $\alpha=2$.
Actually, it is proved for equal-mass planar three-body problem,
and we will prove for general-mass case in this paper.
$^\dag$In this paper, we will extend the proof for Saari's homographic conjecture
to \emph{general-mass} planar three-body problem for $\alpha=1, 2$.
\end{table}
\end{center}

\subsection{Saari's original conjecture}

In 1969, Donald Saari~\cite{SaariOriginal} gave a conjecture;
\begin{conjecture}[Saari's original conjecture, 1969]
Under  Newton potential ($\alpha=1$), if a  motion has constant moment of inertia
then the motion is a relative equilibrium. 
Namely, only possible motion having constant moment of inertia
is a rotation around the centre of mass
as if the $N$-bodies were fixed to a rigid body.
\end{conjecture}

Some
people tried to prove this conjecture more than 30 years
without any positive results.
However, the discovery of the figure-eight solution in 2000
by Chenciner and Montgomery~\cite{C&M} in the three-body problem
under  Newton potential
make us attend to this conjecture because this solution
has almost constant moment of inertia but is not 
a relative equilibrium.

First successful achievement  was made by Christopher McCord~\cite{McCord} 
in 2004.
He proved this conjecture for equal masses case
in planar three-body problem under  Newton potential ($\alpha=1$).
Finally, in  conference ``Saarifest 2005'' held at Guanajuato, Mexico,
Richard Moeckel~\cite{MoeckelSaarifest,MockelProof} proved this conjecture for three-body problem
with general masses
in any spacial dimension
greater than or equal to 2. 

\subsection{Generalised Saari's conjecture}
Saari's original conjecture was generalised
to  homogeneous potentials
given by (\ref{defOfPotential}).

\begin{conjecture}[Generalised Saari's conjecture]
Saari's original conjecture can be extended to 
homogeneous potential with $\alpha \ne -2$ and $2$.
The rectilinear equal mass three-body problem
under $\alpha=-4$ is also excluded.
\end{conjecture}

The harmonic oscillator $\alpha=-2$ is excluded,
because there are trivial counter examples for this potential.
Actually, we can simply construct motions with constant moment of inertia
while each body moves on each ellipse. 
%
For example,
$\bi{q}_k=(a_k \cos(\omega t), b_k \sin(\omega t))$,
$\omega^2 = \sum m_k$
at the centre of mass frame  is a solution of the
equation of motion for $\alpha=-2$.
Then,
the parameters that satisfy
$\sum m_k a_k^2=\sum m_k b_k^2=c=$ constant
makes $I=c$.

In 2006, Gareth E. Roberts~\cite{Roberts} found a counter example 
of this  conjecture in the strong force potential ($\alpha=2$).
The figure-eight solution in the strong force potential ($\alpha=2$)
is also a counter example.
These two motions  have constant moment of inertia but is not 
a relative equilibrium.
This exceptional behaviour of the $N$-body problem in $\alpha=2$
was  already pointed out by Alain Chenciner~\cite{Chenciner1997} in 1997.
Actually, he noticed that the Lagrange-Jacobi identity 
for $\alpha\ne 0$
yields
\begin{equation}
\frac{d^2 I}{dt^2}
=4E+2\left(2\alpha^{-1}-1\right)U.
\end{equation}
Therefore, $I=$ constant makes $U=$ constant for $\alpha\ne 2$,
while $U$ can vary in time for $\alpha=2$.
For $\alpha=2$, integrating $d^2 I/dt^2=4E$,
we get $I=2Et^2+c_1 t+c_2$ with constant parameter $c_1$ and $c_2$.
So, any motion with initial condition $E=0$ and $c_1=0$ has constant moment of inertia.

One more known counter example for generalised Saari's conjecture is
rectilinear motion in the equal mass three-body problem
under the potential $\alpha=-4$ that was described in the section \ref{sec:Intro}.
For this case, $U=-9I^2/8$ is an identity.
Therefore, $\partial U/\partial x_i = -(9I/2) x_i$
makes any motion with constant $I$ a harmonic oscillation, 
%
which is rectilinear not a relative equilibrium rotation
\cite{FFOEvolution}.

\subsection{Saari's homographic conjecture}
In the next day of the same conference where  Moeckel proved
Saari's original conjecture
for three-body problem,
Saari gave a talk and extended his conjecture
in another way~\cite{SaariSaarifest,SaariCollisions},
which is the conjecture \ref{homographicConjecture}.

Saari's homographic conjecture is actually an extension of  original 
and generalised conjecture.
Indeed, for $\alpha\ne 2$, $I=$ constant makes $U=$ constant,
thus makes $\mu=\alpha I^{\alpha/2}U=$ constant.
So, Saari's homographic conjecture contains original conjecture
and generalised conjecture for $\alpha\ne 2$.
%
We expect that this conjecture is true for all $\alpha\ne -2, -4$.

The counter example of Roberts and figure-eight solution both in $\alpha=2$
has constant $I$ and non-constant $U$, therefore non-constant $\mu$.
So, these two examples do not satisfy the assumption of the homographic conjecture. 
Therefore, they are not the counter example for this conjecture.
We expect that  Saari's homographic conjecture is true for $\alpha=2$.
Actually, in this paper, we will prove the conjecture for $\alpha=2$
in planar three-body problem with general masses. 

On the other hands,
the potential for $\alpha=-2, -4$ are really exceptions for Saari's homographic conjecture,
because $\mu$ is identically equal to a constant value in these potentials.

Florin Diacu, Ernesto P\'erez-Chavela, and Manuele Santoprete~\cite{DiacuCollinear}
in 2005 proved this conjecture for collinear $N$-body problem
for any $\alpha$.
Florin Diacu, Toshiaki Fujiwara, Ernesto P\'erez-Chavela
and Manuele Santoprete~\cite{DiacHomographic}
in 2008
showed that the conjecture is true for many set of initial conditions
for planar three-body problem.
In this paper~\cite{DiacHomographic},
the authors  call this conjecture  ``Saari's homographic conjecture''
to distinguish this conjecture from similar two other ``Saari's conjecture''.
The present authors
in 2012
proved the conjecture for planar equal-mass three-body problem
under the strong force potential~\cite{FFOTStrongForce}
and under  Newton potential~\cite{FFOTNewton}.
In this paper, we extend our proof to general masses case.

\section{Dynamical variables and equations of motion}
\label{sec:DynamicalVariables}
\subsection{Notations}
In this paper, we consider the planar three-body problem.
We identify a two dimensional vector $\bi{a}=(a_x,a_y)\in \mathbb{R}^2$
and a complex number $a=a_x+ia_y\in \mathbb{C}$.
Inner and outer products are defined by 
$a\cdot b=a_x b_x+a_y b_y$ and
$a\wedge b=a_x b_y-a_y b_x$.
The partial differentiation by $a$ is defined
\begin{equation}
\frac{\partial}{\partial a}=\frac{\partial}{\partial a_x}+i\frac{\partial}{\partial a_y}.
\end{equation}
For example, for $|a|^2=a_x^2+a_y^2$,
$\partial |a|^2/\partial a=2(a_x+ia_y)=2a$.

\subsection{Dynamical variables}
We take the centre of mass frame. So the position vectors $q_k$ always satisfy
\begin{equation}
m_1q_1+m_2q_2+m_3q_3=0,
\end{equation}
and the moment of inertia is expressed by $I=\sum m_k |q_k|^2$.

According to Richard Moeckel and Richard Montgomery~\cite{M&M},
we define  the ``shape variable'' $\zeta\in \mathbb{C}$
by the ratio of two Jacobi vectors $J_1$ and $J_2$,
\begin{equation}
J_1=q_2-q_1,\,
J_2=q_3-\frac{m_1q_1+m_2q_2}{m_1+m_2}
=\left(\frac{m_1+m_2+m_3}{m_1+m_2}\right)q_3,
\end{equation}
\begin{equation}
\label{defOfZeta}
\zeta
=\frac{J_2}{J_1}
=\left(\frac{m_1+m_2+m_3}{m_1+m_2}\right)
	\left(\frac{q_3}{q_2-q_1}\right).
\end{equation}
\begin{figure}
   \centering
   \includegraphics[width=10cm,bb=0 0 651 321]{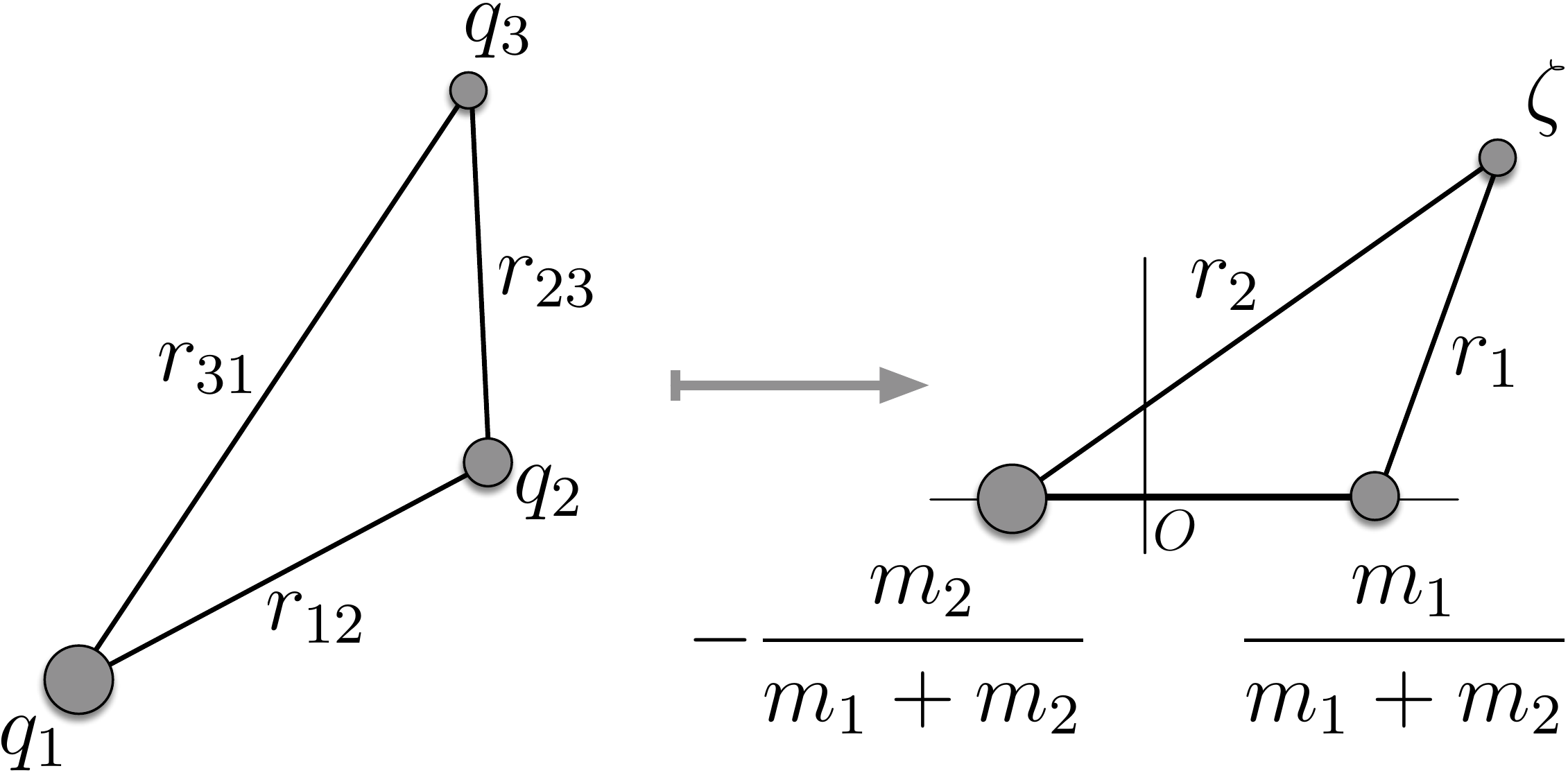} 
   \caption{
   The configuration of $\{q_k\}$ (left)
   and the shape variable $\zeta$ (right).
 A similarity transformation that involves a parallel transformation
   $z \mapsto (z-q_1)/(q_2-q_1)-m_2/(m_1+m_2)$
   maps $q_1$ and $q_2$ to fixed points,
   and $q_3 \mapsto \zeta$.
   The mutual distances $r_{ij}=|q_i-q_j|$ (left) and
   the two-center bipolar coordinates
   $r_1=r_{23}/r_{12}$
   and $r_2=r_{31}/r_{12}$  (right)  are also shown.
   }
   \label{fig:shapeVariables}
\end{figure}
The variable $\zeta$ has a simple geometric interpretation.
Consider a similarity transformation
that involves a parallel transformation
\begin{equation}
z \mapsto \frac{z-q_1}{q_2-q_1}-\frac{m_2}{m_1+m_2}.
\end{equation}
%
The points $q_1$, $q_2$ are mapped to fixed points
\begin{equation}
q_1 \mapsto -\frac{m_2}{m_1+m_2},\,
q_2 \mapsto \frac{m_1}{m_1+m_2},
\end{equation}
then the image of $q_3 \mapsto \zeta$
represents the shape
of the triangle.
See figure \ref{fig:shapeVariables}.
It is convenient to use the following rescaled ``shape variable'' $\eta$
instead of $\zeta$,
\begin{equation}
\label{defOfn}
n=\frac{(m_1+m_2)^2m_3}{(m_1+m_2+m_3)m_1m_2},
\end{equation}
\begin{equation}
\label{defOfEta}
\eta
=\sqrt{n}\,\zeta
=\frac{q_3}{q_2-q_1}
	\sqrt{\frac{(m_1+m_2+m_3)m_3}{m_1m_2}}.
\end{equation}

Let us define $\xi_k=q_k/(q_2-q_1)$
that satisfy $\xi_2-\xi_1=1$ and $m_1\xi_1+m_2\xi_2+m_3\xi_3=0$.
Explicit expression for $\xi_k$ by $\eta$ is
\begin{equation}
\label{defOfXik}
\eqalign{
\xi_1&=-\frac{m_2}{m_1+m_2}
	-\frac{\eta}{m_1+m_2}\, \sqrt{\frac{m_1m_2m_3}{m_1+m_2+m_3}},\\
\xi_2&=\frac{m_1}{m_1+m_2}
	-\frac{\eta}{m_1+m_2}\, \sqrt{\frac{m_1m_2m_3}{m_1+m_2+m_3}},\\
\xi_3&=\eta \sqrt{\frac{m_1m_2}{(m_1+m_2+m_3)m_3}}.
}
\end{equation}
Obviously, the triangles made by $\{q_k\}$ and by $\{\xi_k\}$ are similar
with the common centre of mass.
Therefore, there are $r \ge 0$ and $\phi \in \mathbb{R}$, such that
\begin{equation}
q_k=r e^{i\phi} \frac{\xi_k}{\sqrt{\sum m_\ell |\xi_\ell|^2}}.
\end{equation}
We take the variables $r$, $\phi$ and $\eta$ as the dynamical variables.

The moment of inertia \eref{defOfMomentOfInertia} is given by
\begin{equation}
I=\sum m_k|q_k|^2=r^2.
\end{equation}
The kinetic energy
is expressed by the variables $r$, $\phi$ and $\eta$,
\begin{equation}
\label{theKineticEnergyFinal}
\frac{K}{2}
=\frac{\dot{r}^2}{2}
	+\frac{r^2}{2}\left(
		\dot{\phi}
		+\frac{\eta \wedge \dot{\eta}}
			{1+|\eta|^2}
		\right)^2
	+\frac{r^2}{2}\frac{|\dot{\eta}|^2}{(1+|\eta|^2)^2},
\end{equation}
where dots placed over variables
represent the derivative with respect to  time.
%
The terms of the right-hand side of \eref{theKineticEnergyFinal}
represent
the kinetic energy for the size motion, for the rotation
and for the motion in shape
respectively.
The potential function 
\eref{defOfPotential} for $\alpha\ne 0$
is expressed as
\begin{equation}
\label{potentialFunction}
\eqalign{
\fl
U
=\frac{\mu(\eta)}{\alpha r^\alpha},\\
\fl
\mu(\eta)
=\left(\frac{m_1 m_2}{m_1+m_2}(1+|\eta|^2)\right)^{\alpha/2}\\
	\left(
	m_1m_2
	+\frac{m_2m_3}{\left|m_1/(m_1+m_2)-\eta/\sqrt{n}\right|^\alpha}
	+\frac{m_3m_1}{\left|m_2/(m_1+m_2)+\eta/\sqrt{n}\right|^\alpha}
	\right).
}
\end{equation}
Thus, we obtained expressions for the kinetic energy \eref{theKineticEnergyFinal},
the potential function \eref{potentialFunction},
and thus the Lagrangian $L$ and the total energy $E$
are represented 
by the variables $r$, $\phi$ and $\eta$.

\subsection{The equations of motion}

Since the variable $\phi$ is cyclic,
the angular momentum $C$ is constant of motion.
\begin{equation}
C=\frac{\partial L}{\partial \dot{\phi}}
=r^2 
	\left(
		\dot{\phi}
		+\frac{(\eta\wedge\dot{\eta})}
			{1+|\eta|^2}
		\right)
=\mbox{ constant}.
\end{equation}

The equation of motion for $r$ is
\begin{equation}
\label{ddotr}
\ddot{r}
=\frac{C^2}{r^3}+\frac{r|\dot{\eta}|^2}{(1+|\eta|^2)^2}-\frac{\mu(\eta)}{r^{\alpha+1}}.
\end{equation}
Multiplying both sides of \eref{ddotr} by $\dot{r}$, we obtain
\begin{equation}
\frac{d}{dt}\left(\frac{\dot{r}^2}{2}\right)
=-\frac{d}{dt}\left(\frac{C^2}{2r^2}\right)
	+\frac{|\dot{\eta}|^2}{(1+|\eta|^2)^2}
	\frac{d}{dt}\left(\frac{r^2}{2}\right)
	+\frac{\mu}{\alpha } \frac{d}{dt}\left(\frac{1}{r^\alpha}\right).
\end{equation}
Then, using this equation and the energy conservation $dE/dt=0$, we obtain
the following relation which was first derived by Saari~\cite{SaariCollisions},
\begin{equation}
\frac{d\mu}{dt}
=\frac{\alpha r^{\alpha-2}}{2} \frac{d}{dt}
	\left( r^4 \frac{|\dot{\eta}|^2}{(1+|\eta|^2)^2} \right).
\end{equation}
This equation shows that
the variation in $\mu$ is
proportional to 
the variation in 
the kinetic energy of the shape motion multiplied by $r^2$.
Let us define $v^2$ as
\begin{equation}
v^2 = r^4 \frac{|\dot{\eta}|^2}{(1+|\eta|^2)^2}.
\end{equation}
Then the total energy is given by
\begin{equation}
\label{totalEnergy}
E=\frac{\dot{r}^2}{2}+\frac{C^2+v^2}{2r^2}-\frac{\mu}{\alpha r^\alpha}
\end{equation}
and $v=$ constant if and only if $\mu=$ constant.
Inspired by  Saari's relation,
let us introduce a new ``time'' variable $\tau$ by
\begin{equation}
\label{defOfTau}
\left(\frac{r^2}{1+|\eta|^2}\right)\frac{d}{dt}=\frac{d}{d\tau}.
\end{equation}

The equation of motion for $\eta$
in the time variable $\tau$ is
\begin{equation}
\label{eqOfMotionForEta}
\frac{d^2\eta}{d\tau^2}
=\frac{\displaystyle 2i\left(-C+\eta\wedge\frac{d\eta}{d\tau}\right)}{1+|\eta|^2}\frac{d\eta}{d\tau}
	+\frac{r^{2-\alpha}}{\alpha} \frac{\partial \mu}{\partial\eta}.
\end{equation}

Now, consider a motion that has a constant value of $\mu$.
Then, by Saari's relation, the motion must have constant value of $v^2$.
We have two cases.
\begin{equation}
v^2=\left| \frac{d\eta}{d\tau}\right|^2=
\cases{
0 & (homographic motion),\\
> 0 & (non-homographic motion).
}
\end{equation}

For homographic motion, the equation of motion \eref{eqOfMotionForEta}
demands that
the shape variable must satisfy $\partial \mu/\partial \eta=0$.
We know five solutions:
two Lagrange configurations
and three Euler configurations.

\section{Necessary condition for non-homographic motion}
\label{sec:NecessaryCondition}
Saari's homographic conjecture claims that
non-homographic motion with constant $\mu$
is not realised.
In this section, we assume the existence of 
a non-homographic motion with constant $\mu$,
and will derive a necessary condition for the motion 
to satisfy the equation of motion.

\subsection{Necessary condition in the Cartisian coordinates}
Since such motion satisfy
\begin{equation}
\frac{d\mu}{d\tau}
=\frac{d\eta}{d\tau}\cdot \frac{\partial \mu}{\partial \eta}
=0
\mbox{ and }
\left| \frac{d\eta}{d\tau}\right|^2=v^2 >0,
\end{equation}
the ``velocity'' in the shape variable $d\eta/d\tau$ 
must be orthogonal to the gradient vector $\partial \mu/\partial \eta$
and must have the magnitude $v$.
Therefore, the ``velocity'' is uniquely determined by the gradient vector and $v$,
\begin{equation}
\label{theVelocity}
\frac{d\eta}{d\tau}
=\frac{iv}{|\partial \mu / \partial \eta|}\frac{\partial \mu}{\partial \eta}.
\end{equation}
Here, $v \in \mathbb{R}$ and $v\ne 0$.
In the $\eta$ plane, the motion may pass through a critical point $\partial\mu/\partial\eta=0$.
However, we assume a motion with finite $v$ 
and the critical point is discrete.
Therefore, we can find a part of motion with finite length
where $\partial \mu/\partial \eta \ne 0$.
In the following arguments,
we assume $\partial \mu/\partial \eta \ne 0$ without loss of generality.

Does this motion satisfy the equation of motion?
To give an answer we calculate
the component of the acceleration $d^2\eta/d\tau^2$
in the orthogonal component to the velocity $d\eta/d\tau$,
because parallel component to the velocity is always zero
both in the equation of motion \eref{eqOfMotionForEta} 
and in the motion \eref{theVelocity}.
From the velocity \eref{theVelocity} and its derivative by $\tau$, 
using $d/d\tau=d\eta/d\tau \cdot d/d\eta$, we obtain
\begin{equation}
\label{curvatureByEquipotential}
\frac{d\eta}{d\tau}\wedge\frac{d^2\eta}{d\tau^2}
=\frac{v^3}{(\mu_x^2+\mu_y^2)^{3/2}}
	\left(
	\mu_x^2 \mu_{yy}
	-2\mu_x \mu_y \mu_{xy}
	+\mu_y^2 \mu_{xx}
	\right),
\end{equation}
where $x, y\in \mathbb{R}$ is defined by $\eta=x+iy$
and $\mu_x =\partial\mu/\partial x$, $\mu_y=\partial\mu/\partial y$,
etc....
On the other hand, the equation of motion and 
the velocity \eref{theVelocity} yields
\begin{equation}
\label{curvatureByEquationOfMotion}
\fl
\frac{d\eta}{d\tau}\wedge\frac{d^2\eta}{d\tau^2}
=\frac{2v^2}{1+(x^2+y^2)}\left(
		-C+\frac{v}{\sqrt{\mu_x^2+\mu_y^2}}(x\mu_x+y\mu_y)
	\right)
	-\frac{r^{2-\alpha}v}{\alpha}\sqrt{\mu_x^2+\mu_y^2}.
\end{equation}
Two expressions in \eref{curvatureByEquipotential} and \eref{curvatureByEquationOfMotion} must be the same.
Thus, we get a necessary condition
that must be satisfied by a non-homographic motion
with constant $\mu$,
\begin{eqnarray}
\label{nessesaryCondition}
\fl
\frac{r^{2-\alpha}}{\alpha}
=\frac{-2Cv}{(1+x^2+y^2)\sqrt{\mu_x^2+\mu_y^2}}
	+\frac{2v^2}{(1+x^2+y^2)(\mu_x^2+\mu_y^2)}(x\mu_x+y\mu_y)
	\nonumber\\
	\quad\quad
	-\frac{v^2}{(\mu_x^2+\mu_y^2)^2}
		\left(
			\mu_x^2 \mu_{yy}
			-2\mu_x \mu_y \mu_{xy}
			+\mu_y^2 \mu_{xx}
		\right).
\end{eqnarray}

The right-hand side of the necessary condition
is written in the
Cartesian
coordinate $(x, y)$.
It is convenient to write the right-hand side
in a coordinate free  form.
The kinetic energy for the shape motion
in the equation \eref{theKineticEnergyFinal}
naturally defines the distance squared $ds^2$ and the metric tensor $g_{ij}$
as follows,
\begin{equation}
ds^2
=\frac{dx^2+dy^2}{(1+x^2+y^2)^2}
=g_{ij}dx^i dx^j,
\, 
g_{ij}=\frac{\delta_{ij}}{(1+x^2+y^2)^2}.
\end{equation}
Here 
the repeated indices are understood to be summed.
The vector $(dx^1, dx^2)$ is identified to be $(dx, dy)$ and 
$\delta_{ij}$ represents the Kronecker symbol,
\begin{equation}
\delta_{ij}=\delta^{ij}=
\cases{
1 & for $i=j$,\\
0 & for $i \ne j$.
}
\end{equation}
This metric space is called ``Shape Sphere''.
This sphere is exactly the Riemann sphere 
of the complex plane $x+iy$.
This fact was first noticed by George Lema\^itre~\cite{Lemaitre}
and used by 
Hsiang and Straume~\cite{HsiangWY, Hsiang2006},
Chenciner and Montgomery~\cite{C&M},
Montgomery and Mockel~\cite{M&M},
Kuwabara and Tanikawa~\cite{KuwabaraTanikawa}. 

The inverse and the determinant of the metric are
\begin{equation}
g^{ij}=(1+x^2+y^2)^2\, \delta^{ij},\,
|g|=\det(g_{ij})=\frac{1}{(1+x^2+y^2)^4}.
\end{equation}
Let us define the following three scalars,
\begin{eqnarray}
\fl
|\nabla \mu|^2
=g^{ij}(\partial_i \mu)(\partial_j \mu)
=(1+x^2+y^2)^2 (\mu_x^2+\mu_y^2),
\label{threeScalars1}
\\
\fl
\Delta \mu
=\frac{1}{\sqrt{|g|}}\partial_i
	\left(g^{ij}\sqrt{|g|}\partial_j \mu\right)
=(1+x^2+y^2)^2 (\mu_{xx}+\mu_{yy}),
\label{threeScalars2}
\\
\fl
\lambda
=g^{ij}(\partial_i \mu)\left(\partial_j |\nabla\mu|^2\right)\nonumber\\
=4(1+x^2+y^2)^3 (x\mu_x+y\mu_y)(\mu_x^2+\mu_y^2)\nonumber\\
	\quad\quad
	+2(1+x^2+y^2)^4 (\mu_x^2 \mu_{xx}+2\mu_x\mu_y\mu_{xy}+\mu_y^2\mu_{yy}).
\label{threeScalars3}
\end{eqnarray}
%
In each equality, the first step is definition of each scalar, and the last step is a representation in $(x,y)$ coordinates.
The derivative with respect to time $t$ is given by
\begin{equation}
\label{derivativeByt}
\frac{d}{dt}
=\frac{d\eta}{dt}\cdot\frac{d}{d\eta}
=\frac{(1+x^2+y^2)v}{r^2\sqrt{\mu_x^2+\mu_y^2}}
	\Big( (\partial_x \mu) \partial_y -(\partial_y \mu) \partial_x 
	\Big)
=\frac{v}{r^2 |\nabla\mu|}D.
\end{equation}
Where, $D$ is a differential operator defined by
\begin{equation}
D
=\frac{1}{\sqrt{|g|}}\epsilon^{ij}(\partial_i \mu)\partial_j
=(1+x^2+y^2)^2\Big(
	(\partial_x \mu) \partial_y -(\partial_y \mu) \partial_x 
	\Big),
\end{equation}
and $\epsilon^{ij}$ is 
the L\'evi-Civit\`a anti-symmetric symbol
\begin{equation}
\epsilon^{ij}=
\cases{
1 & for $i=1, j=2$,\\
-1 & for $i=2, j=1$,\\
0 & for $i=j$.
}
\end{equation}

Using these scalars, the necessary condition
\eref{nessesaryCondition} is written
in the coordinate 
free
expression,
\begin{equation}
\label{necessaryCondition}
\frac{r^{2-\alpha}}{\alpha}
=\frac{-2Cv}{|\nabla\mu|}
	+\frac{v^2 \lambda}{2|\nabla\mu|^4}
	-\frac{v^2 \Delta\mu}{|\nabla\mu|^2}.
\end{equation}

\subsection{Necessary condition in  two-center bipolar coordinates}
In this section, we will show a method to rewrite the necessary condition
\eref{necessaryCondition} in the two-center bipolar coordinates
defined by
%
\begin{equation}
\label{defOfTwoCenterBipolarCoordinates}
r_1=\left|\zeta-\frac{m_1}{m_1+m_2}\right|=\frac{|q_2-q_3|}{|q_1-q_2|},\,
r_2=\left|\zeta+\frac{m_2}{m_1+m_2}\right|=\frac{|q_3-q_1|}{|q_1-q_2|}.
\end{equation}
See figure \ref{fig:shapeVariables}.
Although the  coordinates $\eta=x+iy=\sqrt{n}\, \zeta$ are useful to describe the Lagrangian
and to get the equations of motion,
they are not convenient to express the necessary condition.
The expression of the condition in $(x,y)$ coordinates is  lengthy and complex,
while in  $r_1$ and $r_2$ is relatively short and simple.

In the variables $x$ and $y$,
\begin{equation}
r_1^2=\left(\frac{x}{\sqrt{n}}-\frac{m_1}{m_1+m_2}\right)^2+\frac{y^2}{n},\,
r_2^2=\left(\frac{x}{\sqrt{n}}+\frac{m_2}{m_1+m_2}\right)^2+\frac{y^2}{n}.
\end{equation}
Inversely, 
\begin{equation}
\cases{
x=\frac{(m_1-m_2)\sqrt{n}}{2(m_1+m_2)}+\frac{\sqrt{n}}{2}(r_2^2-r_1^2),\\
y=\pm \frac{\sqrt{n}}{2}
	\sqrt{
			\left( 1-(r_1-r_2)^2\right)
			\left((r_1+r_2)^2-1\right)
		}.
}
\end{equation}
Then, distance squared 
$ds^2=(dx^2+dy^2)/(1+x^2+y^2)^2$
is given by
\begin{equation}
\fl
ds^2
=\frac{4m_1m_2m_3(m_1+m_2+m_3)r_1r_2
		\left(r_1r_2(dr_1^2+dr_2^2)-(r_1^2+r_2^2-1)dr_1 dr_2
		\right)}
		{(1-(r_1-r_2)^2)((r_1+r_2)^2-1)(m_1m_2 + m_2m_3r_1^2+m_3m_1r_2^2)^2}.
\end{equation}
Then the metric tensor for this coordinates is defined by
\begin{equation}
\fl
g_{ij}
=\frac{4m_1m_2m_3(m_1+m_2+m_3)r_1r_2}
		{(1-(r_1-r_2)^2)((r_1+r_2)^2-1)(m_1m_2 + m_2m_3r_1^2+m_3m_1r_2^2)^2}	
		\left(\begin{array}{cc}
		a&b\\
		b&a
		\end{array}\right),
\end{equation}
with $a=r_1 r_2$ and $b=-(r_1^2+r_2^2-1)/2$.
The inverse metric $g^{ij}$ and $|g|^{1/2}$ are
\begin{eqnarray}
\fl
g^{ij}
&=\frac{(m_1m_2+m_2m_3r_1^2+m_3m_1r_2^2)^2}{m_1m_2m_3(m_1+m_2+m_3)}
	\left(\begin{array}{cc}
	1&c\\
	c&1
	\end{array}\right),
	\mbox{ with }c=(r_1^2+r_2^2-1)/(2r_1r_2),
\end{eqnarray}
%
\begin{equation}
\fl
|g|^{1/2}
=\frac{2m_1m_2m_3(m_1+m_2+m_3)r_1r_2}
	{(m_1m_2+m_2m_3r_1^2+m_3m_1r_2^2)^2
	\sqrt{(1-(r_1-r_2)^2)((r_1+r_2)^2-1)}
	}.
\end{equation}
Using $\mu$ for $\alpha\ne 0$ expressed as functions of $r_\ell$, $\ell=1,2$,
\begin{equation}
\mu
=\left(
		\frac{m_1m_2+m_2m_3r_1^2+m_3m_1r_2^2}{m_1+m_2+m_3}
	\right)^{\alpha/2}
	\left(
		m_1m_2+\frac{m_2m_3}{r_1^\alpha}+\frac{m_3m_1}{r_2^\alpha}
	\right),
\end{equation}
three scalars \eref{threeScalars1}--\eref{threeScalars3} and 
thus necessary condition \eref{necessaryCondition} are expressed as a function of $r_l$.

\section{Proof of the conjecture}
\label{sec:Proof}
\subsection{Proof for the strong force potential}
For the strong force potential $\alpha=2$,
the left-hand side of the necessary condition \eref{necessaryCondition}
is constant
and the right-hand side is a function of $r_\ell^2$,
\begin{equation}
\label{necessaryConditionStrongForce}
\frac{1}{2}
=\frac{-2Cv}{|\nabla\mu|}
	+\frac{v^2 \lambda}{2|\nabla\mu|^4}
	-\frac{v^2 \Delta\mu}{|\nabla\mu|^2}.
\end{equation}

\begin{figure}
   \centering
   \includegraphics[width=8cm]{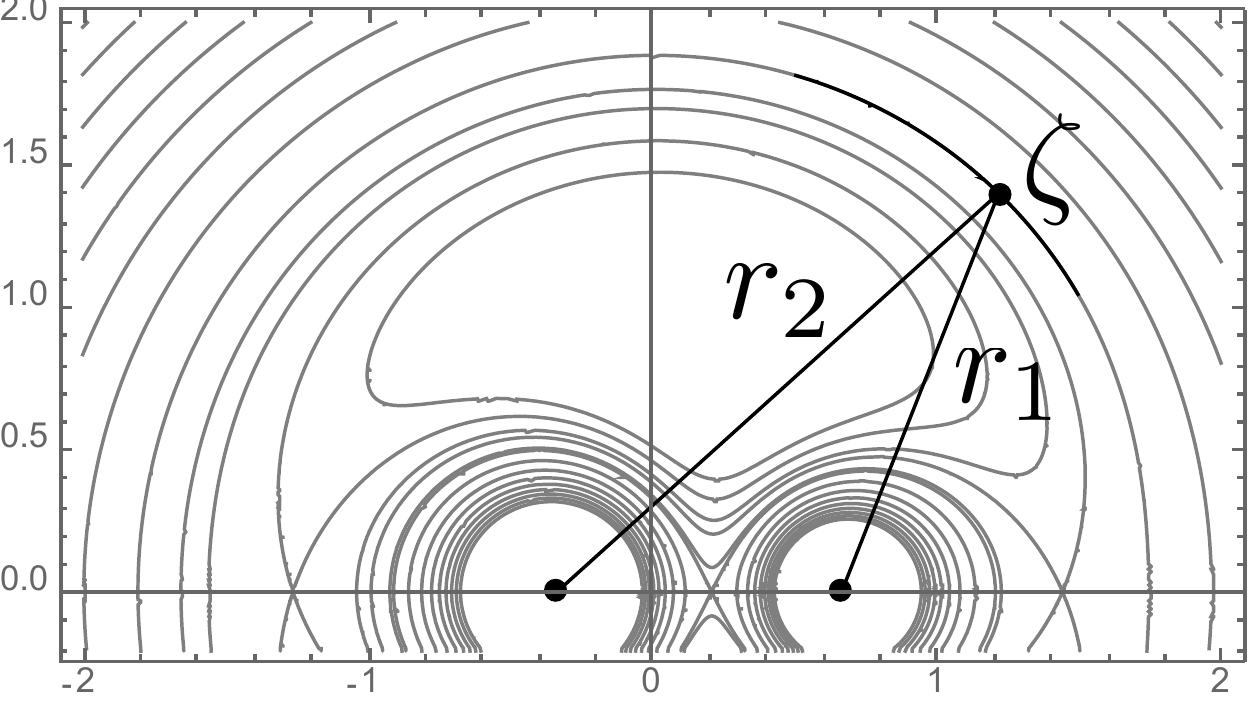} 
   \caption{
Contours of $\mu=$ constant for $\alpha=2$.
   If a non-homographic motion having constant $\mu$ exist,
   the shape variable $\zeta$ moves on one of the contours.
   }
   \label{fig:contoursOfMu}
\end{figure}

Two variables $r_1^2$ and $r_2^2$ are not independent,
because we are considering a motion
that has $\mu(r_1^2, r_2^2)=$ constant.
We have only one independent variable.
See figure \ref{fig:contoursOfMu}.
%
A possible choice of one independent variable is, say, $r_1^2$.
Solving $\mu(r_1^2, r_2^2)=\mu$ for $r_2^2$,
we will obtain $r_2^2=r_2^2(m_k, \mu, r_1^2)$.
Then, the necessary condition \eref{necessaryConditionStrongForce}
will be in the form
$1/2=F(m_k, C, v, \mu, r_1^2)$.
This is a condition for independent variable $r_1^2$
with constants $m_k, C, v, \mu$.
However, this choice breaks the invariance of 
the condition \eref{necessaryConditionStrongForce}
under the simultaneous exchange of
$m_1 \leftrightarrow m_2$ and $r_1^2 \leftrightarrow r_2^2$.
Breaking this symmetry will make our analysis
complex.
%
Let us write a desirable variables $\{\nu, \rho\}$
that would keep this symmetry,
easy to solve the variable change
$\{r_1^2, r_2^2\} \leftrightarrow \{\nu, \rho\}$,
and simple to eliminate one variable using $\mu=$ constant.

Our choice for $\{\nu, \rho\}$ is
\begin{equation}
\label{defOfNuandRhoForStrongForce}
\cases{
\nu
=m_1m_2+m_2m_3r_1^2+m_3m_1r_2^2,\\
\rho
=m_1m_2+\frac{m_2m_3}{r_1^2}+\frac{m_3m_1}{r_2^2}.
}
\end{equation}
Obviously, these variables keep the symmetry.
The equation \eref{defOfNuandRhoForStrongForce}
is easy to solve for $\{r_1^2, r_2^2\}$
because this equation is quadratic.
Moreover, we simply eliminate $\nu$ by
\begin{equation}
\nu =\frac{\tilde{\mu}}{\rho},
\mbox{ where }
\tilde{\mu}=(m_1+m_2+m_3)\mu.
\end{equation}

We have two solutions of $r_\ell^2=r_\ell^2(\tilde{\mu},\rho)$
for the equation \eref{defOfNuandRhoForStrongForce}.
Substituting a solution
into the necessary condition \eref{necessaryConditionStrongForce},
we obtain a necessary condition for $\rho$
as follows, 
\begin{equation}
\label{necessaryConditionStrongForce2}
\frac{1}{2}=F(C, v^2, m_k, \tilde{\mu}, \rho).
\end{equation}
Now,
if there is a non-homographic motion with constant $\mu=\tilde{\mu}/(m_1+m_2+m_3)$,
there is some finite physical  interval of $\rho$
where the condition \eref{necessaryConditionStrongForce2}
is satisfied. See figure \ref{fig:contoursOfMu}.
Since the right-hand side of the condition \eref{necessaryConditionStrongForce2}
is an analytic function of $\rho$,
this condition must be satisfied
for whole complex plane of $\rho \in \mathbb{C}$.
Therefore, the condition \eref{necessaryConditionStrongForce2}
must be satisfied near the origin of $\rho$,
although this region is unphysical.

Two solutions of \eref{defOfNuandRhoForStrongForce} are
\begin{equation}
\label{sol1ForString}
\fl
r_1^2=-\frac{m_3}{m_1}+\frac{m_3(m_1^2m_3^2-\tilde{\mu})}{m_1^2m_2}\left(\frac{\rho}{\tilde{\mu}}\right)+O(\rho^2),
\, 
r_2^2=\frac{\tilde{\mu}}{m_1m_3\rho}+\frac{m_2(m_3^2-m_1^2)}{m_1^2m_3}+O(\rho),
\end{equation}
and
\begin{equation}
\label{sol2ForString}
\fl
r_1^2=\frac{\tilde{\mu}}{m_2m_3\rho}+\frac{m_1(m_3^2-m_2^2)}{m_2^2m_3}+O(\rho),
\,
r_2^2=-\frac{m_3}{m_2}+\frac{m_3(m_2^2m_3^2-\tilde{\mu})}{m_1m_2^2}\left(\frac{\rho}{\tilde{\mu}}\right)+O(\rho^2).
\end{equation}
The latter is given by 
simultaneous exchange of $r_1 \leftrightarrow r_2$
and $m_1 \leftrightarrow m_2$
in the former.
Substituting  the  solution  \eref{sol1ForString}
into the condition \eref{necessaryConditionStrongForce},
we obtain
%
%
\begin{eqnarray}
\label{theConditionForAlphaEq2}
\fl
\frac{1}{2}
=\frac{(m_1+m_2+m_3)v}{2 m_1^2 m_2^2}
\Bigg\{
-v \left(\tilde{\mu}+m_1^2 \left(m_2^2+m_2 m_3-m_3^2\right)
+m_1 m_2 m_3^2+m_2^2 m_3^2\right)\nonumber\\
+2 i C m_1 m_3 \sqrt{m_2^3 (m_1+m_2+m_3)}
\Bigg\}
\left(\frac{\rho}{\tilde{\mu}}\right)^2
+O(\rho^3).
\end{eqnarray}
Each of three terms in the right-hand side of \eref{necessaryConditionStrongForce}
contributes to $O(\rho^2)$.
Note that there is no term of order $\rho^0$ in the right-hand side
of \eref{theConditionForAlphaEq2}
while the left-hand side is $1/2$.
Therefore, this condition cannot be satisfied
by the solution \eref{sol1ForString}.
For the solution \eref{sol2ForString},
we have similar result.
Only the difference from the equation \eref{theConditionForAlphaEq2}
is the exchange of $m_1$ and $m_2$.
Thus,
the condition \eref{necessaryConditionStrongForce}
cannot be satisfied.
Namely, there is no non-homographic motion with constant $\mu$.
This is a proof of Saari's homographic conjecture
for the strong force potential $\alpha=2$.

\subsection{Proof for Newton potential}
For Newton potential $\alpha=1$,
the necessary condition \eref{necessaryCondition}
\begin{equation}
\label{necessaryConditionForNewton}
r
=\frac{-2Cv}{|\nabla\mu|}
	+\frac{v^2 \lambda}{2|\nabla\mu|^4}
	-\frac{v^2 \Delta\mu}{|\nabla\mu|^2}
\end{equation}
determines the size variable in the form $r=r(C,v,m_k,r_\ell)$.
Then by the equation \eref{derivativeByt},
$\dot{r}$ is also given in the form $\dot{r}=\dot{r}(C,v,m_k,r_\ell)$.
Thus the total energy \eref{totalEnergy} is  written in the form
$E=E(C,v,m_k,r_\ell)$.

For Newton potential, let us take new variables
\begin{equation}
\label{defOfNuRhoForNewton}
\cases{
\nu
=m_1m_2+m_2m_3r_1^2+m_3m_1r_2^2,\\
\rho=m_1m_2+\frac{m_2m_3}{r_1}+\frac{m_3m_1}{r_2}.
}
\end{equation}
Then, we eliminate $\nu$ by
\begin{equation}
\nu = \left( \frac{\tilde{\mu}}{\rho}\right)^2,
\mbox{ where }
\tilde{\mu} = \mu \sqrt{m_1+m_2+m_3}.
\end{equation}
The equation \eref{defOfNuRhoForNewton} is 
a quartic equation for $r_\ell$.
Let one of the solutions be $r_\ell=r_\ell(m_k, \tilde{\mu},\rho)$.
Substituting this solution into the expression of $E$,
we will obtain total energy in the form
\begin{equation}
\label{totalEnergyInRho}
E=E(C,v,m_k,\tilde{\mu},\rho).
\end{equation}

Let us assume that there is a physical value of 
$C, v, m_k, \tilde{\mu}$
and finite physical interval of $\rho$
where the right-hand side of equation \eref{totalEnergyInRho} is constant. 
For physical region, $\tilde{\mu}$ is always greater than $(m_1 m_3)^{3/2}$
and $(m_2 m_3)^{3/2}$.
This is because
\begin{equation}
\tilde{\mu}
> \sqrt{m_3m_1r_2^2}
	\left(
		\frac{m_3m_1}{r_2}
	\right)
=(m_1m_3)^{3/2},
\end{equation}
and similar inequality $\tilde{\mu}>(m_2 m_3)^{3/2}$.
Since the right-hand side of equation \eref{totalEnergyInRho}
is an analytic function of $\rho$,
the right-hand side must be constant
for whole region of the complex plane $\rho \in \mathbb{C}$.
Therefore, the expression \eref{totalEnergyInRho} must be constant
near the origin of $\rho$
for some physical value of $C$, $v$, $m_k>0$,
 and $\tilde{\mu}>(m_1 m_3)^{3/2}, (m_2 m_3)^{3/2}$.

The four solutions of \eref{defOfNuRhoForNewton} are
\begin{equation}
\label{seriesForr1andr2}
\fl
\cases{
r_1=-\frac{m_3}{m_1}\left(
	1
	+\frac{\tilde{\mu}\pm(m_1m_3)^{3/2}}{m_1m_2\tilde{\mu}}\rho
	+O(\rho^2)
	\right),\\
r_2=\frac{1}{\sqrt{m_1m_3}}\left(
	\mp\frac{\tilde{\mu}}{\rho}
	+O(\rho)
	\right),
}
\end{equation}
and simultaneous exchange 
of $r_1 \leftrightarrow r_2$ and $m_1 \leftrightarrow m_2$.
Then three quantities in the necessary condition 
for the solutions in \eref{seriesForr1andr2} are
\begin{eqnarray}
\frac{1}{|\nabla \mu|^2}
=\frac{(m_1+m_2+m_3)^2 m_3^3}{m_1^3 m_2}\left(\frac{\rho}{\tilde{\mu}}\right)^6
	+O(\rho^7),\label{nablaMu1}\\
\lambda
=\frac{3m_1^4(\mp(m_1 m_3)^{3/2}-\tilde{\mu})}{2m_3^6 (m_1+m_2+m_3)^{7/2}}
	\left(\frac{\tilde{\mu}}{\rho}\right)^{10}
	+O(1/\rho^{9}),\label{lambda1}\\
\Delta\mu
=\frac{(m_1+m_2)m_3^3-m_1^3(m_2+m_3)}{m_3^3(m_1+m_2+m_3)^{3/2}}
	\left(\frac{\tilde{\mu}}{\rho}\right)^3
	+O(1/\rho^2).\label{deltaMu1}
\end{eqnarray}
Therefore,
the dominant term in the necessary condition \eref{necessaryConditionForNewton}
near the origin of $\rho$
is $v^2 \lambda/(2|\nabla\mu|^4)$.
Thus, the condition \eref{necessaryConditionForNewton} yields
\begin{equation}
\label{theConditionForAlphaEq1}
r=\mp\frac{3v^2((m_1m_3)^{3/2}\pm\tilde{\mu})\sqrt{m_1+m_2+m_3}}
		{4m_1^2 m_2^2}
		\left(\frac{\rho}{\tilde{\mu}}\right)^2
		+O(\rho^3).
\end{equation}
Then
\begin{equation}
\frac{1}{r^2}
=\frac{16 m_1^4 m_2^4 }
	{9(m_1+m_2+m_3)v^4((m_1 m_3)^{3/2}\pm \tilde{\mu})^2}
	\left(\frac{\tilde{\mu}}{\rho}\right)^4
	+O(\rho^{-3}).
\end{equation}
And using the equation \eref{derivativeByt},
we obtain
\begin{equation}
\fl
\left(\frac{\rho}{\tilde{\mu}^2} \frac{d\rho}{dt}\right)^2
=- \frac{64 m_1^9 m_2^7 }
	{81m_3^3 (m_1+m_2+m_3)^{3} v^6 ((m_1 m_3)^{3/2}\pm \tilde{\mu})^4}
	\left(\frac{\tilde{\mu}}{\rho}\right)^8
	+O(\rho^{-7}),
\end{equation}
Therefore, near the origin of $\rho$,
the dominant term in the total energy \eref{totalEnergy}
is the kinetic term for size motion $\dot{r}^2/2$.
We obtain
\begin{equation}
E
=-\frac{16m_1^5 m_2^3 }
{9m_3^3(m_1+m_2+m_3)^2((m_1m_3)^{3/2}\pm\tilde{\mu})^2v^2}
\left(\frac{\tilde{\mu}}{\rho}\right)^8
+O(\rho^{-7}).
\end{equation}
The other two solutions of $r_\ell$ give the total energy in
exchange of $m_1 \leftrightarrow m_2$.
Note that the coefficient of the term $(\tilde{\mu}/\rho)^8$
is not zero.

Thus the total energy $E$ cannot be constant
near the origin of $\rho$.
This means that there is no non-homographic motion
with constant $\mu$.
This is a proof of Saari's homographic conjecture.

\section{Summary and discussions}
We proved Saari's homographic conjecture
for planar three-body problem
under Newton potential ($\alpha=1$) and 
the strong force potential ($\alpha=2$)
for general masses.

To describe the motion in shape,
we used the shape variable 
$\zeta \in \mathbb{C}$ in the equation \eref{defOfZeta} 
or $\eta \in \mathbb{C}$ in the equation \eref{defOfEta} 
introduced by Moeckel and Montgomery.
We wrote the Lagrangian in
the size variable $r$, rotation variable $\phi$
and the shape variable $\eta$.
The equations of motion for these variables
were given.

Then,
we assumed the existence of a non-homographic motion
that has constant configurational measure $\mu$.
This motion must satisfy the necessary condition
\eref{necessaryCondition}.
Finally, we showed that
any non-homographic motion with constant $\mu$
are not able to satisfy the necessary condition.
This is our proof.

In the final stage of our proof,
we changed the variables $\eta\in \mathbb{C}$
to two-center bipolar coordinates $(r_1, r_2)$
defined in the equation 
\eref{defOfTwoCenterBipolarCoordinates},
then to $(\tilde{\mu}, \rho)$
in \eref{defOfNuandRhoForStrongForce}
or \eref{defOfNuRhoForNewton}.
The variables $(\tilde{\mu}, \rho)$ is useful to prove
Saari's homographic conjecture.
Because we assume $\mu=\tilde{\mu}/(m_1+m_2+m_3)^{\alpha/2}=$ constant,
the only one free variable is $\rho$.
This choice of the variables makes our proof simple.

We have two comments for the variable $(\tilde{\mu}, \rho)$.
One is an alternative method to calculate
$|\nabla \mu|$, $\Delta\mu$ and $\lambda$.
In this paper, we expressed these quantities in
the variables $(r_1, r_2)$.
Then, put $r_k =r_k(\tilde{\mu},\rho)$ to get $|\nabla \mu|$ etc...
in a series of $\rho$.
An alternative method is direct calculation of
them using the metric in $(\tilde{\mu},\rho)$ space,
$ds^2=G_{ij}dx^i dx^j$ and $(dx^1, dx^2)=(d\tilde{\mu}, d\rho)$.
Here, we write the metric in $(\tilde{\mu},\rho)$ space  $G_{ij}$.
This is simply given by the variable change
from  $(dr_1, dr_2)$ to $(d\tilde{\mu}, d\rho)$.
Then, we will get the metric $G_{ij}(\tilde{\mu},\rho)$
in a series of $\rho$.
Using this metric, we directly calculated 
$|\nabla\mu|^2=G^{ij}(\partial_i\mu)(\partial_j\mu)$ etc...
and got the same results in equations 
\eref{theConditionForAlphaEq2} and
\eref{theConditionForAlphaEq1}.

Another comment is a difficulty to extend our method
to general $\alpha$,
for example, to $\alpha=\sqrt{2}$.
According to this paper, a naive choice of $(\nu, \rho)$ will be
\begin{eqnarray}
\nu=m_1 m_2 +m_2 m_3 r_1^2+ m_3 m_1 r_2^2,\\
\rho=m_1 m_2 +m_2 m_3/r_1^\alpha+ m_3 m_1/r_2^\alpha,
\end{eqnarray}
and
\begin{equation}
\tilde{\mu}
=(m_1+m_2+m_3)^{\alpha/2}\mu
=\nu^{\alpha/2}\rho.
\end{equation}
However, it will be difficult to solve this equations to get
$r_1$ and $r_2$ in a power series of $\rho$.
It would be better to find another  variables.

\ack
This research of one of the author T.~Fujiwara  has been  supported by
Grand-in-Aid for Scientific Research 23540249 JSPS.

\end{document}